# Simulated dose painting of hypoxic sub-volumes in pancreatic cancer stereotactic body radiotherapy

Ahmed M. Elamir[1,2], Teodor Stanescu[1,2], Andrea Shessel[1], Tony Tadic[1,2], Ivan Yeung[1,2,3], Daniel Letourneau[1,2], John Kim[1,2], Jelena Lukovic[1,2], Laura A. Dawson[1,2], Rebecca Wong[1,2], Aisling Barry[1,2], James Brierley[1,2], Steven Gallinger[4,5], Jennifer Knox[6,7], Grainne O'Kane[4,6,7], Neesha Dhani[6,7], Ali Hosni[1,2], Edward Taylor[1,2]

[1]Radiation Medicine Program, Princess Margaret Cancer Centre, Toronto, Canada.
[2]Department of Radiation Oncology, University of Toronto, Toronto, Canada.
[3]Stronach Regional Cancer Centre, Southlake Regional Health Centre, Newmarket, Canada.
[4]Ontario Institute for Cancer Research, PanCuRx Translational Research Initiative, Toronto, Canada
[5]Department of Surgery, University of Toronto, Toronto, Canada
[6]Division of Medical Oncology and Hematology, Princess Margaret Cancer Center, Toronto, Canada.
[7]Department of Medicine, University of Toronto, Toronto, Canada.

## Abstract

Dose painting of hypoxic tumour sub-volumes using positron-emission tomography (PET) has been shown to improve tumour control *in silico* in several sites, predominantly head and neck and lung cancers. Pancreatic cancer presents a more stringent challenge, given its proximity to critical gastro-intestinal (GI) organs-at-risk (OARs), anatomic motion, and impediments to reliable PET hypoxia quantification. A radiobiological model was developed to estimate clonogen survival fraction (SF), using $^{18}$F-fluoroazomycin arabinoside PET (FAZA PET) images from ten patients with unresectable pancreatic ductal adenocarcinoma to quantify oxygen enhancement effects. For each patient, four simulated five-fraction stereotactic body radiotherapy (SBRT) plans were generated: 1) a standard SBRT plan aiming to cover the planning target volume with 40 Gy, 2) dose painting plans delivering escalated doses to a maximum of three FAZA-avid hypoxic sub-volumes, 3) dose painting plans with simulated spacer separating the duodenum and pancreatic head, and 4), plans with integrated boosts to geometric contractions of the gross tumour volume (GTV). All plans saturated at least one OAR dose limit. SF was calculated for each plan and sensitivity of SF to simulated hypoxia quantification errors was evaluated. Dose painting resulted in a 55% reduction in SF as compared to standard SBRT; 78% with spacer. Integrated boosts to hypoxia-blind geometric contractions resulted in a 41% reduction in SF. The reduction in SF for dose-painting plans persisted for all hypoxia quantification parameters studied, including registration and rigid motion errors that resulted in shifts and rotations of the GTV and hypoxic sub-volumes by as much as 1 cm and 10 degrees. Although proximity to OARs ultimately limited dose escalation, with estimated SFs (~$10^{-5}$) well above levels required to completely ablate a ~10 cm$^3$ tumour, dose painting robustly reduced clonogen survival when accounting for expected treatment and imaging uncertainties and thus, may improve local response and associated morbidity.

# Introduction

The role of radiation therapy (RT) in the treatment of locally advanced pancreatic cancer remains controversial, with a number of studies reporting minimal to no added efficacy of conventionally fractionated RT beyond standard-of-care chemotherapy [1-4]. Over the past decade, hypo-fractionated stereotactic body radiation therapy (SBRT) has increasingly been incorporated into the management of pancreatic cancer since it enables the delivery of a higher biologically effective dose (BED) [5-7] and may be easier to coordinate with systemic chemotherapy. It has been hypothesized that the higher BEDs may result in increased local response, the lack of which is a major impediment to the management of this disease [8-13]. Such dose escalation is blind to the effects of hypoxia, the state of low oxygenation that often arises in solid tumours, including pancreatic cancer [14, 15]. Hypoxia downwardly modifies the BED by as much as a factor of three [16-18], providing a rationale for patient-specific dose escalation. Confounding the ability to dose-escalate pancreatic tumours, however, are their proximity to critical gastrointestinal organs-at-risk (OARs), especially the duodenum.

As an illustration of the challenges to dose escalation in pancreatic cancers, radiobiological modelling of pancreatic tumour response has demonstrated that at least 80 $Gy_{10}$ BED is needed to achieve at least partial local radiologic response in 50% of cases [19]. For five-fraction SBRT, this translates to a dose of approximately 43 Gy, exceeding typical dose limits for duodenum, bowel, and stomach by 7-10 Gy [20-23]. Given the proximity of pancreatic tumours to the duodenum, this means that dose distributions in pancreatic tumours for hypo-fractionated, high-BED treatments are inevitably highly heterogeneous, often undercovering regions of the planning target volume (PTV), or even the gross tumour volume (GTV) [9, 10]. At the same time, hypoxia in pancreatic tumours is heterogeneous, between patients [14, 15, 24], and intra-tumourally [24]. This leads us to the central question posed by this manuscript: can simultaneous integrated boosts to hypoxic sub-volumes in pancreatic tumours potentially improve local response, by pushing dose into hypoxic regions while maintain clinical goals for OARs?

Such "dose painting" to hypoxic tumour sub-volumes has shown promise as a way to overcome the de-sensitizing effects of hypoxia [25-29], delivering higher doses to regions that are determined to be hypoxic using e.g., positron-emission tomography (PET) imaging of hypoxia-sensitive agents. *In silico* feasibility and clinical studies of PET-based dose painting have largely focussed on head and neck [30-35] and lung tumours [36-39]. Pancreatic cancer challenges the dose painting paradigm in a number of ways: first, the aforementioned proximity to critical gastro-intestinal OARs. Second, its apparent greater innate radioresistance, possibly arising from aberrant DNA damage sensing and repair pathways, cell-cycle checkpoint activation, and modifications to autophagy and apoptosis mechanisms [40]. Third, transport impediments to mid-size molecular agents that reduce sensitivity of PET hypoxia imaging of pancreatic tumours [41]. Finally, inter- and intra-fraction anatomic motion and deformation [42-45], which impacts the reliability of image registration between hypoxia PET images and the primary treatment planning images.

In this manuscript, we simulated the feasibility and efficacy of RT dose painting in patients with locally advanced, unresectable pancreatic cancer who underwent [18]F-fluoroazomycin arabinoside (FAZA) PET imaging for hypoxia. Five-fraction SBRT plans were generated for each patient using several strategies—standard SBRT with the goal of covering the PTV with 40 Gy, dose-painting based on FAZA PET images, and boosts to geometric contractions of the GTV [9, 10]. The relative efficacy of the plans was assessed by estimating the clonogen cell survival fraction (SF) using a modified linear-quadratic formalism that accounted for oxygenation levels, as determined from FAZA PET images. Sensitivity to several uncertainties in quantifying hypoxia, including noise, transport impediments, registration, and motion errors, was quantified by recalculating SF for a given optimized radiotherapy plan, having transformed the FAZA uptake map to simulate these errors.

## Methods

### Study design and research ethics board approval

This was a single institute retrospective study to assess the dosimetric feasibility and efficacy of SBRT with dose escalation to hypoxic sub-volumes defined by FAZA PET. The study was initiated after approval from our institutional research ethics board.

### Patient data

Imaging data for ten patients with unresectable pancreatic adenocarcinoma (PDAC) enrolled in a previous research ethics board approved hypoxia imaging study [15] were used to retrospectively study RT planning strategies to overcome hypoxia-induced radioresistance. Of the twenty-three patients from the original study with available images, sixteen were unresectable; of these, ten exhibited appreciable levels of hypoxia (hypoxic fraction greater than 3%; see below for definition) and were included in the present analysis.

### Imaging data and region of interest contours

For each patient, SBRT plans were generated on computed tomography (CT) images rigidly co-registered to FAZA PET images acquired sequentially on the same scanner. The details of image acquisition were previously reported [15]. Diagnostic contrast-enhanced CT scans were used to assist delineation of GTV. Following standard PET-based hypoxia quantification methodology [46, 47], hypoxia was quantified in each tumour voxel by the ratio ("TBR", the tissue-to-blood ratio) of FAZA activity in that voxel to the mean activity in the abdominal aorta at the level of the pancreas, contoured as a 1 cm diameter region-of-interest to minimize partial volume effects [48]. PET voxels were $3.9 \times 3.9 \times 3.9$ mm$^3$ and CT voxels were 1.0 mm$^2$ in the axial plane and 2.0 mm longitudinally.

Internal target volumes (ITVs) were generated from GTVs using a 3 mm expansion in all directions except inferiorly, where a 6 mm expansion was used. This margin corresponds to the average breathing displacements of GTVs assessed from a motion study of ten pancreatic cancer patients with abdominal compression at our institution using 4D-CT (data not published), also consistent with published data [44]. ITVs were expanded by 5 mm to create the primary PTV (PTV_4000), as per our institutional practice. Throughout this manuscript, numerical suffixes in PTV names indicate goal dose levels in cGy. Hypoxic boost volumes GTV_h1.2, GTV_h1.3, and GTV_h1.4 were defined as regions in the GTV for which TBR exceeded 1.2, 1.3, and 1.4, respectively, corresponding to a range of standard thresholds for identification of hypoxic regions [15, 46, 49]. Boost volumes < 0.5 cm$^3$ were not contoured. Boost PTV volumes, PTV_h1.2_4500, PTV_h1.3_5000, and PTV_h1.4_5400, were generated by uniformly expanding the hypoxic boost GTV volumes by 5 mm in all directions, consistent with the margin applied to the primary GTV; no internal margin was used since the PET acquisition was continuous for ~20 minutes and thus was assumed to incorporate the effects of breathing motion. For the geometric boost strategy [9, 10] (see below for details), a high-dose PTV (PTV_HD_5000) was contoured for each patient as the GTV excluding 5 mm planning-at-risk volumes (PRVs), as well as an ultra-high dose PTV (PTV_UHD_5400), equal to PTV_HD uniformly contracted by 1 cm. Justification for these goal prescription doses is given below.

The OARs were contoured at least 2 cm above and below the level of PTV_4000 in the cranial-caudal direction, and included the spinal cord, esophagus, stomach, duodenum, small and large bowel, liver and kidneys. PRVs were created as 3 mm and 5 mm expansions of critical OARs (stomach, duodenum, small bowel, large bowel, and spinal cord). Evaluation structures (Eval_PTV_x, where x indicates target dose) for the primary and hypoxic boost volume PTVs were generated by subtraction of critical OARs (stomach, duodenum, small bowel, large bowel) from the corresponding PTVs. To simulate the effect of adding a hydrogel spacer between the pancreas and duodenum [50], a structure of width 5 mm was contoured between the duodenum and GTV and a modified duodenum contour was created by subtracting this structure from the original duodenum region-of-interest (ROI).

**SBRT planning protocols**

For each patient, four five-fraction SBRT plans were generated in the RayStation treatment planning system (RayStation v6, RaySearch Laboratories, Stockholm, Sweden) using a 2.5 mm grid size and collapsed cone convolution-superposition algorithm: 1) a standard SBRT plan optimized to deliver a total of 40Gy to ≥ 95% of the Eval_PTV_4000; 2) a dose painting plan delivering integrated boosts to hypoxic sub-volumes; 3) a dose painting plan using a modified duodenum ROI to simulate duodenal spacer; and 4) a plan to deliver integrated boosts to geometric sub-volumes of the GTV (PTV_HD_5000, PTV_UHD_5400). This last approach ("geometric boost") replicates that of Crane and colleagues [9, 10]. The rationale for using this geometric boost is that it does not rely on expensive PET imaging and should have similar efficacy if hypoxic sub-volumes are localized in the center of the tumour, as often assumed. Target goals for all plans are shown in Table 1. All plans were optimized to satisfy the global constraint D1cm$^3$ < 5850 cGy and saturate at least one of the critical OAR constraints shown in Table 2 and hence, were nominally isotoxic. The prescriptions of 45, 50, and 54 Gy for hypoxic boost volumes were chosen to minimize cancer cell survival theoretically, for a fixed mean dose, subject to our OAR constraints; see below. To be consistent with these dose levels, for the geometric boost plans, we chose 50 and 54 Gy for the HD and UHD sub-volume prescription. In terms of BED$_{10}$ (biologically effective dose using α/β= 10 Gy), these correspond to 100 and 112 Gy$_{10}$, respectively. For comparison, Crane and colleagues proposed delivering approximately 100 and 150 Gy$_{10}$ to the same volumes, in 15 or 25 fractions [9, 10].

**Radiobiological modelling of SBRT plans using FAZA PET imaging**

Survival fractions for all simulated SBRT plans were calculated using the linear-quadratic (LQ) model expression for cell survival, modified to account for oxygen enhancement ratio (OER) effects, with oxygenation determined from FAZA uptake data. This last step required simulation of spatial oxygen distributions for modelled capillary patterns to determine oxygen partial pressure $p$ probability distribution functions $f(p; v_c)$ for given capillary densities $v_c$. These distributions were then convolved with the partial-pressure dependent LQ model—calculated for a range of doses $D$—and a pharmacokinetic FAZA binding model to determine dose-survival fraction (SF) curves and FAZA TBR values for each $v_c$ simulated. These functions were interpolated and combined to give the survival fraction as a function of dose and FAZA TBR. This workflow is shown schematically in Figure 1; details are given below.

Oxygen partial pressure distributions were calculated by solving the oxygen reaction-diffusion equation [51]

$$D_{O_2}\vec{\nabla}^2 p(\vec{r}) = \frac{c_{max} p(\vec{r})}{k + p(\vec{r})} \tag{1}$$

for the oxygen partial pressure $p(\vec{r})$ at position $\vec{r}$ on a two-dimensional square domain of width $l = 1$ mm, subject to the boundary condition $p(\vec{r}) = p_c$ at the surface of capillaries of radius $r_c = 5$ μm. $D_{O_2}$ is the diffusivity of oxygen, $c_{max}$ is the maximum metabolism, and $k$ is the partial pressure at which metabolism reaches half its maximum value; values for these quantities are shown in Table 3 and are the same as those used by Dasu *et al.* [51] and Petit *et al.* [52]. Equation (1) was solved using a finite-element technique, having randomly distributed the capillaries over the 1 mm² domains for a given capillary areal density $v_c$ ($= N_{cap} \pi (r_c/l)^2$, where $N_{cap}$ is the number of capillaries). Following the approach of Ref. [52], the calculation was repeated at least 100 times for different random capillary placements. The unit-normalized (i.e., $\int_0^\infty dp\, f(p; v_c) = 1$) oxygen partial pressure distribution $f(p; v_c)$ for each value of $v_c$ was then obtained by sampling the data from all simulations for that $v_c$ value. This process was repeated for a range of capillary densities between 0.00125 to 0.025.

The oxygen partial pressure distributions were then combined with a pharmacokinetic model of FAZA uptake in pancreatic tumours [41, 48] and a Michaelis-Menten-type approximation for the binding kinetics of nitroimidazole-based hypoxia PET tracers (such as FAZA) derived by Casciari and colleagues [53] to give the FAZA TBR as a function of capillary density:

$$\text{TBR}(v_c) = A + B \int_0^\infty dp\, f(p; v_c) \frac{p_1}{p + p_1} \tag{2}$$

Here, $p_1 = 1.15$ mmHg [54, 55] is the partial pressure at which binding achieves half its maximum value. $A = 1$ and $B = 1.4$ are dimensionless parameters related to the pharmacokinetic properties of FAZA in pancreatic tumours. A derivation of Equation (2) is given in the Supplementary Materials.

The survival fraction in a voxel with capillary density $v_c$ receiving a dose $D$ was found from the linear-quadratic (LQ) model, modified to account for OER effects [18, 52]:

$$\text{SF}(D, v_c) = \int_0^\infty dp\, f(p; v_c) e^{-\alpha \cdot \text{BED}(p, D)} \tag{3}$$

Here, α is the usual LQ model radiosensitivity parameter; see Table 3. For a total dose $D$ accumulated over $N_f = 5$ fractions, the biologically effective dose is [18, 52]:

$$\text{BED}(p, D) = \frac{D}{\text{OER}(p)} \left(1 + \frac{D}{N_f (\alpha/\beta) \text{OER}(p)}\right) \tag{4}$$

We use α/β = 10 Gy, as usually assumed for early-reacting tissues [56]. The OER—defined as the ratio of the doses under hypoxic (i.e., for an oxygen partial pressure $p$) and completely oxygenated conditions needed to achieve the same level of cell killing—was assumed to depend on the oxygen partial pressure as [18]

$$\text{OER}(p) = \frac{\text{OER}_{\max}(K_m + p)}{K_m + p \cdot \text{OER}_{\max}} \quad (5)$$

where $\text{OER}_{\max} = 3$ is the maximum OER (i.e., when $p = 0$) and $K_m = 3.28$ mmHg [52] is the partial pressure at which the OER achieve half its maximum value. In attributing a single $f(p; v_c)$ to all fractions, we made the conservative approximation that hypoxia does not change during treatment, neglecting the impact of reoxygenation [57] and transient/acute hypoxia [58].

For each simulated capillary density, Equation (3) was calculated for doses $D$ between 0 and 120 Gy, in 1 Gy increments. $\text{TBR}(v_c)$ and $\text{SF}(D, v_c)$ were then interpolated (with respect to $v_c$ and $D$); TBR was inverted to give $v_c = v_c(\text{TBR})$ and combined with $\text{SF}(D, v_c)$, yielding the survival fraction $\text{SF}(D, \text{TBR})$ as a function of dose and FAZA TBR, shown in Figure 2. With this function, we were then able to calculate the SF for each SBRT plans as follows: for each voxel (enumerated by the index $i$), the FAZA uptake $\text{TBR}_i$ was combined with the dose $D_i$ to yield the SF in that voxel. The total tumour SF was then obtained by taking the average of the SF in each voxel:

$$\text{SF} = \frac{1}{N_v} \sum_{i=1}^{N_v} \text{SF}(D_i, \text{TBR}_i) \quad (6)$$

Here, $N_v$ is the number of voxels in the GTV. To accommodate the discrepancy in size between PET voxels and the dose grid, in evaluating Equation (6), PET voxels were re-sized and activity levels linearly interpolated to match the isotropic 2.5 mm voxels used in the dose calculation.

By applying our radiobiological model to the GTV and TBR values therein, Equation (6) neglects the impact of inter- and intra-fraction tumour and hypoxic sub-volume motion. Also, Equation (6) implicitly assumes that there are no setup errors and the dose planned for a given GTV voxel is the dose delivered. The impact of motion, setup errors, and other hypoxia quantification uncertainties will be simulated as described below.

Equation (6) was used to estimate the clonogen survival fraction for the four different planning strategies as well as their sensitivity to hypoxia quantification and geometric errors; we also compared the results with target dose-volume histogram (DVH) parameters shown in Table 1 to determine if a simple DVH criterion could act as a surrogate for plan efficacy (i.e., SF).

**Dose painting prescription levels**

For a fixed mean dose $\bar{D}$ to the GTV, passing into the continuum limit $D_i \rightarrow D(\vec{r})$, the dose distribution $D(\vec{r})$ that minimizes the survival fraction is given by the solution of the Euler-Lagrange equation $\partial \text{SF}/\partial D(\vec{r}) = \mu$, where µ is a Lagrange multiplier that fixes the mean dose $\bar{D} = \int d\vec{r}\, D(\vec{r})/(\text{GTV volume})$ [59]. Fixed mean dose is often taken as a surrogate for iso-toxicity [59-61]. Here, even though all plans are actually iso-toxic (OAR dose-limited), assuming fixed mean dose to determine prescription levels provided a useful starting point for optimization. Further assuming constant TBR values of 1.1, 1.25, 1.35, and 1.45 for the GTV sub-volumes corresponding to TBR<1.2, 1.2<TBR<1.3, 1.3<TBR<1.4, and TBR>1.4, respectively, for a $\bar{D}$ consistent with the oxic region (TBR<1.2) of the GTV receiving a minimum dose of 30 Gy (as per the minimum allowed GTV D99% in Table 1), the doses to GTV_h1.2, GTV_h1.3, GTV_h1.4 that minimize SF are 45 Gy, 50 Gy, and 54 Gy, respectively (having numerically evaluated the Euler-Lagrange equation for the SF shown in Figure 2). These are the prescription levels shown in Table 1. We note that choosing the more ambitious minimum GTV dose of 40 Gy (our nominal prescription dose), SF is minimized by prescribing 67 Gy, 72 Gy, and 76 Gy to the hypoxic sub-volumes. These far exceed what is achievable given our OAR dose constraints.

**Sensitivity analysis for hypoxia quantification errors**

Four types of hypoxia quantification errors were simulated to test the sensitivity of the dose painting plans: 1) random image noise, 2) reference tissue activity errors, 3) FAZA transport effects, and 4) registration errors between the PET and planning image data sets. The last also simulates the effect of rigid motion of the GTV as well as setup errors; deformations of the GTV are expected to be smaller than rigid motions [43-45] and were not simulated. All errors were simulated by an appropriate transformation of the spatial TBR distribution, schematically represented as TBR→TBR'; specific transformations are given below. The effect of these errors was quantified by the ratio $SF_{dose\,painting}$ (TBR')/$SF_{standard}$ (TBR') of the survival fractions calculated for dose painting and standard SBRT plans, evaluated using the transformed TBR. As long as this ratio remains less than one, the dose painting plan results in greater cell-killing than the non-boost, standard SBRT plan.

1. <u>Random image noise and hypoxia fluctuations.</u> Test/re-test measurements of hypoxia PET images have revealed a high degree of repeatability for images acquired within one week, for a range of nitroimidazole-based tracers [62, 63], including for pancreatic cancers [64]. Nonetheless, differences between repeat PET scans remain, resulting from intrinsic PET imaging noise and possibly acute hypoxia. To simulate this, random noise was added to the voxel-scale TBR data set:

$$\text{TBR}_i \rightarrow \text{TBR}_i + \delta_i \qquad (6)$$

$\delta_i$ was randomly sampled from a Gaussian distribution of mean zero and standard deviation σ. $SF_{dose\,painting}/SF_{standard}$ was evaluated for a range of σ values between 0 and 0.7 and results expressed in terms of the Pearson correlation coefficient between the transformed and original TBR datasets.

2. <u>Reference tissue activity errors.</u> Variability in the imaged activity level of blood in the abdominal aorta used to determine TBR can arise from PET partial volume and liver spill-over effects [48]. To simulate these, the TBR dataset was scaled linearly as

$$\text{TBR}_i \rightarrow \lambda \text{TBR}_i, \qquad (7)$$

where λ represents the ratio between the image-derived blood activity level and the true value (e.g., measured activity in drawn blood). $SF_{dose\,painting}/SF_{standard}$ was evaluated for λ spanning 0.8 to 1.2, encompassing the expected range of reference tissue activity variance [48, 65].

3. <u>FAZA transport effects.</u> Pancreatic cancer hypo-perfusion and regions of slow molecular diffusivity can result in reduced distribution volumes for molecular agents such as FAZA [41]. Expecting that voxels with low TBR values are more affected by transport impediments, we used the following transformation to simulate transport effects:

$$\text{TBR}_i \rightarrow \text{TBR}_i + \gamma(\text{TBR}_{max} - \text{TBR}_i) \qquad (8)$$

Here, $\text{TBR}_{max}$ is the maximum TBR value in the GTV and γ is a parameter varied between 0 and 0.5, encompassing the range of FAZA distribution volumes in pancreatic cancer [41]. $\text{SF}_{\text{dose painting}}/\text{SF}_{\text{standard}}$ was evaluated after application of this mapping for different γ values. Results were presented as a function of the change in hypoxic fraction due to the simulated transport effects, allowing for comparison with the results of [41].

4. <u>Rigid registration errors between PET and planning image sets and motion errors.</u> Both translational and rotational registration errors were simulated by spatial shifts and rotations of the TBR "map" (i.e., the three-dimensional spatial distribution)

$$\text{TBR}_i \rightarrow \text{TBR}_{i+\Delta,} \qquad (9)$$

where Δ represents either a translation or rotational coordinate shift and, as before, *i* enumerates the voxel. For each boost plan, translational shifts and rotations with respect to all three axes were randomly sampled at least 500 times from the range -20 to 20 mm and -10 to 10 degrees, in increments consistent with our dose grid size of 2.5 mm. Rotations of the TBR map were carried out with respect to the centroid of each GTV. $\text{SF}_{\text{dose painting}}/\text{SF}_{\text{standard}}$ was calculated and results presented as functions of the amplitude of the translational displacement (i.e., $\sqrt{(\Delta_x)^2 + (\Delta_y)^2 + (\Delta_z)^2}$ for random shifts $\Delta_i$ in the *i*th direction) and the Euler angles corresponding to the three randomly-sampled rotation angles about the x,y, and z axes. Equation (9) also simulated the "worst-case" error due to rigid target motion, since it represents a single displacement and/or rotation, with no possibility of averaging to a smaller value as radiotherapy fractions are delivered. For simplicity, we did not investigate the combined effects of rotation and translation. At most, the errors would add in quadrature.

## Results

### Patients and tumor characteristics

Ten patients who met the inclusion criteria were included in the analysis. Mean GTV volume was 50.4 cm$^3$ (range: 24.4 – 81.8 cm$^3$). Mean hypoxic fraction (fraction of GTV voxels for which TBR>1.2) was 0.20 (range: 0.09 – 0.38). The primary dose-limiting OAR was the duodenum. Mean minimum Euclidean distance (MED) between the GTV and duodenum was 1 mm (range: 0 – 4 mm); mean MED between the lowest hypoxic sub-volume GTV_h1.2 and duodenum was 3 mm (range: 0 – 9 mm). Geometric characteristics of tumors are reported in Table 4.

### Dosimetric characteristics of SBRT plans

Dose painting of hypoxic sub-volumes to ≥ 45 Gy was feasible but challenged by the proximity of OARs such as duodenum. The mean Eval_PTV_h1.2_4500 V45Gy for the dose painting plans was 78% (range: 61%-98%; *p* = 0.09 vs. standard SBRT), compared to 65% (range: 36%-95%) for standard SBRT plans and 75% (range: 57%-99%; *p* = 0.19 vs. standard SBRT; *p* = 0.62 vs. dose painting) for geometric boost plans. The difference between the dose painting and geometric boost plans was more pronounced at the next highest dose level: mean Eval_PTV_h1.3_5000

V50Gy was 69% (range: 41%-99%) and 59% (range: 27%-92%) for the dose painting and geometric boost plans, respectively; $p = 0.31$.

All plans saturated at least one of the OAR dose constraints in Table 2 to within 10 cGy. Duodenum was the most common dose-limiting OAR, with all patient plans being limited either by D0cm$^3$ < 36 Gy (n=1), D0.5cm$^3$ < 36 Gy for the PRV (n=7), or both (n=2). This included the single patient whose GTV was in the tail of the pancreas, and for whom dose was limited by the ascending part of the duodenum. Two patients were dose-limited by small bowel in addition to duodenum. The addition of simulated spacer led to an increase in hypoxic sub-volume coverage for the dose painting plans: mean Eval_PTV_h1.2_4500 V45Gy and Eval_PTV_h1.3_5000 V50Gy increased to 89% (range: 74%-99%) and 85% (range: 67%-100%). Target DVH statistics are summarized in Table 5.

**Clonogen cell survival fractions**

Dose painting resulted in a decreased clonogen cell SF: mean SF values were $5.8 \times 10^{-5}$ (standard SBRT), $3.4 \times 10^{-5}$ (geometric boost; $p = 0.14$ vs standard SBRT), and $2.6 \times 10^{-5}$ (dose painting; $p = 0.08$ vs standard SBRT); see box-and-whisker plot in Figure 3. Dose painting thus reduced SF by ~55% as compared to conventional SBRT plans. In comparison, geometric boost plans resulted in a ~41% reduction in SF as compared to conventional SBRT plans. SF for the dose painting plans was further halved (reduced by ~78% as compared to standard SBRT) by the addition of simulated spacer, to a mean value of $1.3 \times 10^{-5}$; $p = 0.12$ vs boost without spacer; $p = 0.01$ vs standard SBRT). For all target objectives shown in Table 1, SF values evaluated for all forty plans were most strongly correlated with Eval_PTV_h1.2_4500 V45Gy ($r = -0.74$); see right panel of Figure 3. In comparison, Eval_PTV_4000 V40Gy and PTV_UHD_5400 V54Gy were only modestly predictive of SF with $r$ values of -0.44 and -0.55, respectively. Coverage of the higher-level hypoxic sub-volumes was less strongly predictive of cell survival than coverage of Eval_PTV_h1.2: correlations between Eval_PTV_h1.3_5000 V50Gy and Eval_PTV_h1.4_5400 V54Gy with SF were -0.68 and -0.53. PTV_HD_5000 V50Gy did not predict SF ($r = -0.30$).

**Sensitivity of dose painting plans to hypoxia quantification errors**

The reduction in cell survival due to dose painting as compared to standard SBRT plans, quantified by the ratio SF$_{dose\ painting}$/SF$_{standard}$, persisted for all simulated hypoxia quantification errors (Figure 4) and PET-to-planning image set registration errors (Figure 5). Namely, SF$_{dose\ painting}$/SF$_{standard}$ remained less than one for all simulated errors, including rotation and translation errors as large as 10° and 1 cm.

## Discussion

In contrast to the present study, most *in silico* dose painting studies have quantified efficacy in terms of the tumour control probability (TCP), the probability of killing *all $N_c$* clonogens in a tumour (i.e., complete local control) and approximated by [26, 60, 66]

$$\text{TCP} \approx e^{-N_c \text{SF}} \tag{10}$$

For a 50 cm$^3$ tumour—the average value of the tumours in the present study—, assuming a clonogen density between 10$^6$ and 10$^8$ cm$^{-3}$ [26, 60], the SF would need to be between 10$^{-8}$ and 10$^{-10}$ to have a 50% chance of ablating the tumour (TCP = 0.5). Although dose painting robustly (i.e., accounting for errors likely to arise in the clinical implementation of dose painting) reduced SF in our study as compared to the standard SBRT plans, the resulting 55% reduction, from SF ≈ 5.8×10$^{-5}$ to 2.6×10$^{-5}$ is still not nearly enough to completely ablate the tumours studied, and TCP was essentially zero for all simulated RT plans, a result of the inability to deliver high enough doses to tumours that were uniformly in close proximity to duodenum and other critical gastro-intestinal organs-at-risk.

Re-evaluating the survival fractions for the dose painting plans under simulated oxic conditions (OER = 1), the mean SF value was 1.9×10$^{-8}$, expected to result in a nonzero TCP. Thus, even with the limitations imposed on the GTV dose by surrounding OARs, the dose plans used in this study may have been sufficient to achieve complete eradication of the pancreatic tumours under completely oxic conditions. Hypoxia was thus the primary cause of the simulated high cancer clonogen survival fractions and corresponding treatment failure.

The estimated cell survival magnitudes are sensitive to the choice of parameters used, especially the intrinsic radiosensitivity α under oxic conditions. Most theoretical dose painting studies have used $\alpha = 0.35 - 0.41$ Gy$^{-1}$ [31, 35, 38, 39, 52, 60], values primarily derived from in vitro studies of head and neck cancers [67], but also consistent with in vivo studies of lung cancer data [68]. In three human pancreatic cancer cell lines, El Shafie [58] *et al.* [69] found a mean α of 0.25 Gy$^{-1}$ (range: 0.19 – 0.29 Gy$^{-1}$) [69], the value we used in this study. While the absolute cell survival is highly sensitive to the choice of α, the relative survival fraction between treatment plans (i.e., the ratio of SF for two plans) is not and so we expect dose painting plans to remain superior to standard SBRT ones, irrespective of the choice of α.

Our results for cell survival neglected changes in hypoxia that likely take place throughout treatment [57, 58]. Reoxygenation during the short duration of SBRT is possible [57], although is probably not of sufficient magnitude to appreciably impact our results. Further work is needed to quantify the impact of acute or "cyclic" hypoxia [58], although we note that dose painting resulted in a robust reduction in cancer clonogen survival fraction when accounting for random noise (Figure 4(a.)). Such noise may simulate cyclic hypoxia to the extent that it models transient fluctuations in tumour blood flow, and hence, FAZA uptake. Dose painting remained superior to non-dose painting plans since the fluctuations occurred on top of a constant background of chronic hypoxia. Transient fluctuations in oxygen would reduce estimated survival fraction, however, since the most hypoxic cells at a given fraction are more likely to experience an increase in oxygenation at subsequent fractions.

In addition to assuming constant oxygenation throughout treatment, our modelling also did not directly simulate the effects of inter- and intra-fraction motion, which would likely result in a reduction in the efficacy of dose painting. Randomly displacing and rotating the FAZA spatial uptake map with respect to the plan dose distributions, however, we found that the survival fractions for dose painting plans remained consistently below those for the non-dose painting plans. To the extent that such motion represents a worst-case scenario for anatomic motion, we conclude that the dose painting plans are likely robust against inter- and intra-fraction motion, arising from registration errors or anatomic changes. This was primarily due to the fact that the image-derived hypoxic sub-volumes were typically several cm's in spatial extent, larger than expected motion errors.

We also note that our study patient population was small (10 patients) and may not be representative of the SBRT patients treated at other institutions. For instance, Reyngold and collaborators report reserving SBRT for patients with gastrointestinal structures more than 1 cm from the GTV [9]. In contrast, for all our patients, the duodenum was within 4 mm of the GTV.

Even though dose painting did not appreciably elevate TCP, given the high rate of distant failure in patients with pancreatic cancer [70] and the morbidity associated with local failure [71], we hypothesize that even a modest reduction in clonogen survival may improve local response and clinical outcomes in the first few years following treatment. A number of studies have looked at the impact of dose escalation in pancreatic tumours on overall (OS) and loco-regional recurrence-free (LRRFS) survival [8-13]. A common strategy is to dichotomize patients into high- and low-$BED_{10}$ groups for Kaplan-Meier analysis, where $BED_{10}$ is defined as $D[1 + D/N_f(\alpha/\beta)]$, with $\alpha/\beta = 10$ Gy, and "high" $BED_{10}$ corresponds to $BED_{10} > 70$ $Gy_{10}$ [11-13]. In a retrospective study of 200 patients with locally advanced pancreatic cancer, Krishnan and colleagues found OS at 3 years was 31% and 9%, respectively, in the high- and low-$BED_{10}$ groups, while LRRFS at 1 year was 21% and 9% in these groups [11]. Mean $BED_{10}$ in the two groups was 60 and 83 $Gy_{10}$ and a crude estimate of the dose-response gradient is thus $\gamma \sim 1\%/Gy_{10}$ and $\sim 0.5\%/Gy_{10}$ for OS and LRRFS, respectively. Using the linear quadratic model to define a "mean" $BED_{10}$ as $\overline{BED}_{10} = -\ln SF/\alpha$ in conjunction with the results of our survival fraction analysis, the difference in SF between the standard SBRT and dose-painting plans amounts to an 8% increase in $\overline{BED}_{10}$. For our standard SBRT plans, the mean GTV dose was ~52 Gy, corresponding to a $BED_{10} = 107$ $Gy_{10}$. Taking this as a baseline, the reduction in SF due to dose-painting thus corresponds to an approximately 9 $Gy_{10}$ increase in $BED_{10}$ and hence, based on the results of Krishnan *et al.* [11], we anticipate a ~9% increase in OS at 3 years and ~5% increase in LRRFS at 1 year.

Given the expense of hypoxia PET imaging—both in terms of the imaging agent cost and burden on patients during potentially long imaging sessions—it is noteworthy that a boost to concentric contractions of the GTV[10] performed better than the standard SBRT plans with a mean SF of $3.4 \times 10^{-5}$, only 30% greater than the mean PET dose-painting value. For patients for whom hypoxia was centrally located (e.g., Figure 6(c.)), dose painting offered essentially no improvement beyond the geometric boost plans. Hypoxia was not always centrally located within the GTV, however (e.g., Figure 6(b.)), and for these patients, PET-based dose painting reduced cell survival by as much as a factor of 0.6 as compared to the geometric boost plans.

The addition of simulated gel duodenal spacer resulted in a further reduction in clonogen survival, especially amongst patients for whom the GTV abutted the duodenum. For two patients for whom the minimum Euclidean distance between GTV and duodenum was greater than 2 mm, the spacer offered no added benefit and $SF_{dose\ painting\ with\ spacer}/SF_{dose\ painting\ without\ spacer}$ was 1.01 and 0.99. For the remaining eight patients for whom the minimum distance was less than 1mm, the mean $SF_{dose\ painting\ with\ spacer}/SF_{dose\ painting\ without\ spacer}$ was 0.5 (range: 0.21-0.76). An example case with $SF_{dose\ painting\ with\ spacer}/SF_{dose\ painting\ without\ spacer} = 0.21$ is shown in Figure 7.

The impact of spacer on SF suggests that a further, comparable, reduction in SF can be achieved by motion management strategies that allow for a reduction in the ITV margins [44].

Because the PET images used in the study were acquired continuously and hence, were blurred by abdominal motion, we were unable to simulate the impact of motion reduction in a simple way. However, given that breathing-induced displacement amplitudes of pancreatic tumours (~5 mm-1cm [44]) are on the order of the spacer gel thickness, we expect that e.g., gated delivery of SBRT [23] should result in a comparable decrease in SF as the use of spacer.

Dose painting plans remained superior (smaller SF) as compared to standard SBRT plans when simulating the effects of hypoxia quantification errors, geometric image registration errors, and motion. To the best of our knowledge, this is the first time that such a sensitivity analysis has been undertaken. Although the PET and planning CT images sets used in this study were acquired on the same scanner, minimizing registration errors, the robustness of the dose painting plans to registration errors as large as 1 cm and 10° suggests that patients can be imaged on a PET scanner for hypoxia quantification and then treated under MR guidance for GTV and OAR localization, using a daily adapted planning approach [23, 72]. In addition to the aforementioned size of the hypoxic subvolumes, the robustness of dose painting plans to such errors may have been due in part to our conservative planning approach, requiring GTV D99% to be strictly greater than 30 Gy for all plans, prioritizing primary target coverage over coverage of hypoxic sub-volumes. Had we prioritized hypoxic sub-volume coverage at the expense of primary target coverage, it is likely that the dose painting plans would not be as robust. Dose painting plans also remained superior to standard SBRT ones when accounting for errors in hypoxia quantification arising from the choice of reference tissue as well as imaging agent transport effects [41]. Even though the latter resulted in changes to the shapes of the hypoxic sub-volumes, these were not substantial enough to eliminate the efficacy of dose painting for error values expected to arise clinically [41].

Further adding to the feasibility of dose painting for pancreatic tumours in the clinical setting is the fact that a simple target dose-volume histogram (DVH) metric, the fraction of the evaluation PTV corresponding to the lowest hypoxic sub-level (Eval_PTV_h1.2_4500) receiving 45 Gy was strongly predictive of clonogen survival fraction. This means that plan quality can be assessed using a DVH criterion, without the need for radiobiological modelling at the time of planning. This is particularly important for "online" adaptive treatments using magnetic resonance image guidance [13, 72, 73], where plan adaptation based on daily anatomy must be accomplished quickly.

Although dose painting using standard OAR dose constraints was shown to be feasible and hypothesized to result in a modest improvement in short-term survival, more innovative solutions are needed to overcome hypoxia-induced radioresistance in pancreatic cancer in order to achieve complete tumour ablation. High linear-energy transfer (LET) radiation such as carbon ions can mitigate the impact of hypoxia through a reduction in the OER; clinical trials investigating their efficacy in treating locally advanced pancreatic cancer are underway at a number of institutions [74]. Heavy ion accelerators are expensive, however, and lower-cost solutions are desirable. Hypoxia-activated cytotoxic prodrugs such as evofosfamide have shown promise, improving pancreatic tumour control after radiotherapy in pre-clinical settings [75, 76], although these have yet to see widespread clinical use. Given the predicted three to five order-of-magnitude increase in cell killing needed to achieve complete local ablation of pancreatic tumours, it also likely that a more dramatic improvement is needed. The success of SBRT in treating early-stage lung cancer with doses on the order of 100-120 $Gy_{10}$ BED [77], and the fact that these tumours exhibit comparable levels of hypoxia as pancreatic ones based on PET imaging [78], suggests that increased dose escalation (> 120 $Gy_{10}$) combined with agents that protect gastrointestinal organs for radiation damage [79, 80] may be the best route forward.

## Conclusions

SBRT with dose painting of hypoxic sub-volumes of pancreatic cancer is feasible and robust to expected treatment and imaging errors, but ultimately limited by the proximity of tumours to the duodenum. Using a model to convert $^{18}$F-fluoroazomycin arabinoside (FAZA) PET images of pancreatic tumours to oxygen-enhancement ratio maps, clonogen survival fractions calculated from the linear-quadratic model were reduced by dose painting as compared to standard radiotherapy plans that only sought to optimize planning target volume coverage. Although this reduction was not nearly enough to kill all clonogens in a typical pancreatic tumour, applying our modelling to existing dose escalation study results, we hypothesize that dose painting may improve local response and survival, several years post-treatment.

| Planning approach | Primary target goals | Secondary (boost) target goals |
|---|---|---|
| Non-boost | Eval_PTV_4000: V40 Gy > 90%<br>GTV: D99% > 33 (± 3) Gy | |
| Dose-painting (PET boost) | Eval_PTV_4000: V40Gy > 90%<br>GTV: D99% > 33 (± 3) Gy | Eval_PTV_h1.2_4500: V45Gy > 90%<br>Eval_PTV_h1.3_5000: V50Gy > 90%<br>Eval_PTV_h1.4_5400: V54Gy > 90% |
| Geometric boost | Eval_PTV_4000: V40Gy > 90%<br>GTV: D99% > 33 (± 3) Gy | PTV_HD_5000: V50Gy > 90%<br>PTV_UHD_5400: V54Gy > 90% |

Table 1. Target planning goals for the three simulated planning strategies. Goals for the dose-painting plan with simulated spacer were identical to those shown above. For all plans, the maximum dose to 1cm$^3$ of tumour was kept below 58.5 Gy. Coverage of the Eval_PTV_4000 and GTV was prioritized over boost goals. Parentheses indicate allowed variation for the GTV D99%.

| Structure | | Criterion | Dose Limit |
|---|---|---|---|
| Luminal structures | Stomach | D0cm$^3$ | < 36 Gy |
| | Duodenum | D0.5cm$^3$ | < 35 Gy |
| | Small bowel | D1cm$^3$ | < 33 Gy |
| | Large bowel | D0.5cm$^3$ to corresponding 3mm PRV | < 36 Gy |
| Spinal cord | | D1cm$^3$ | < 15 Gy |
| Esophagus | | D0.5cm$^3$ | < 32 Gy |
| Liver | | D50% | < 12Gy |
| Kidneys | | D75% | < 12 Gy |
| | | Mean Dose | < 10 Gy |

Table 2. OAR dose constraints.

| Parameter | Symbol | Value(s) | Reference |
|---|---|---|---|
| Intrinsic radiosensitivity | $\alpha$ | 0.25[a] $Gy^{-1}$ | [69] |
| Alpha-beta ratio | $\alpha/\beta$ | 10 Gy | - |
| Oxygen partial pressure at which the OER is half its maximum | $K_m$ | 3.28 mmHg | [18] |
| maximum OER value | $OER_{max}$ | 3 | [18] |
| Oxygen partial pressure at which tracer binding is half its maximum | $p_1$ | 1.15[b] mmHg | [54, 55] |
| FAZA pancreas pharmacokinetic parameters | $A$ | 1[c] | [41, 53] |
| | $B$ | 1.4[c] | |
| Oxygen partial pressure at which metabolism is half its maximum | $k$ | 2.5 mmHg | [52] |
| Capillary oxygen partial pressure | $p_c$ | 35 mmHg | [52] |
| Oxygen diffusivity | $D_{O_2}$ | 2000 µm²/s | [52] |
| Maximum oxygen metabolic rate | $c_{max}$ | 15 mmHg/s | [52] |

Table 3. Radiobiological parameters used to determine cell survival fraction (SF) from dose and uptake of FAZA. [a]Average value from three pancreatic cell lines studied in Ref. [69]. [b]Mid-point of the range of values reported in Refs. [54, 55]. [c]See Supplementary Materials for the derivation of these parameters.

| GTV volume [cm³] | GTV_h1.2 volume [cm³] | GTV_h1.3 volume [cm³] | GTV_h1.4 volume [cm³] |
|---|---|---|---|
| 50.4 (range: 24.4-81.8) | 9.9 (range:2.1-19.8) | 3.8 (range:0.2-7.4) | 1.1 (range: 0-3.1) |

Table 4. Geometric characteristics of the ten pancreatic tumours in this study. All 10 patients had contourable GTVh_1.2 and GTVh_1.3 volumes; 1 patient did not have a GTV_h1.4 volume. Volumes < 0.5 cm³ were not contoured.

| Target | DVH metric | Standard SBRT plan mean (SD, p) | Dose painting plan mean (SD, p) | Dose painting plan with spacer mean (SD, p) | Geometric boost mean (SD, p) |
|---|---|---|---|---|---|
| Eval_PTV_4000 | V40Gy | 79% (9, --) | 76% (12, 0.60) | 83% (8, 0.27) | 80% (8, 0.79) |
| GTV | D99% | 3364 cGy (345, --) | 3410 cGy (448, 0.70) | 3781 cGy (434, **0.03**) | 3454 cGy (463, 0.54) |
| Eval_PTV_h1.2_4500 | V45Gy | 65% (19, --) | 78% (14, 0.09) | 89% (8, **0.001**) | 75% (14, 0.19) |
| Eval_PTV_h1.3_5000 | V50Gy | 35% (29, --) | 69% (21, **0.01**) | 85% (12, **0.00**) | 59% (23, **0.05**) |
| Eval_PTV_h1.4_5400 | V54Gy | 7% (10, --) | 52% (29, **0.00**) | 69% (20, **0.00**) | 34% (25, **0.01**) |
| PTV_HD_5000 | V50Gy | | | | 78% (12) |
| PTV_UHD_5400 | V54Gy | | | | 99% (2) |

Table 5. Target dose volume histogram (DVH) statistics for the four planning strategies employed in this study. Standard deviation (SD; in %) and *p* values relative to standard SBRT plan are shown in parentheses.

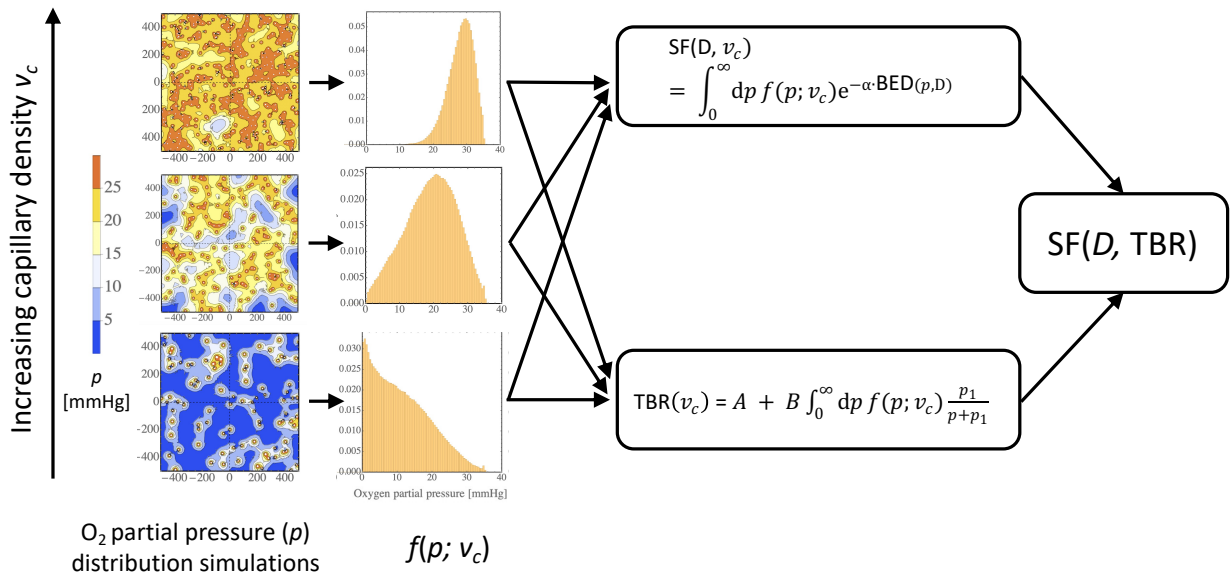

Figure 1. Schematic representation of the calculation of survival fraction (SF) as a function of dose *D* and FAZA uptake (TBR). Left: oxygen partial pressure maps were simulated for randomly chosen capillary placements for specified capillary densities $v_c$ (Equation 1). These calculations were repeated at least 100 times and sampled to give partial pressure probability distributions $f(p;v_c)$. These distributions were convolved with the oxygen partial pressure dependent FAZA binding (tissue-to-blood ratio TBR; Equation 2) and linear quadratic (for SF; Equation 3) models. Finally, these were combined to give the survival fraction SF(*D*,TBR) as a function of dose and FAZA TBR.

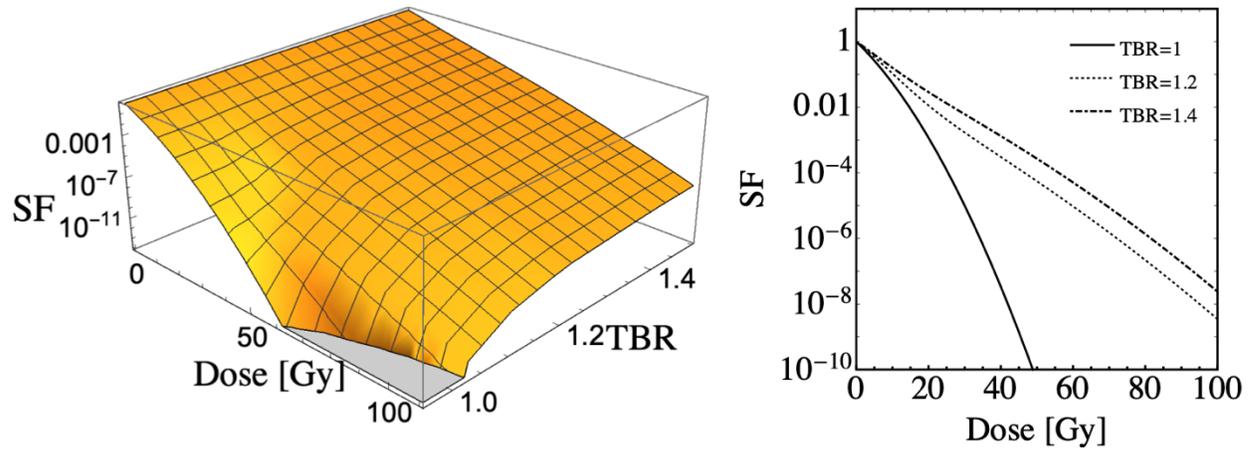

Figure 2. Radiobiological model used to calculate survival fractions. Left: Survival fraction (SF) versus dose and the hypoxia-sensitive FAZA tissue-to-blood ratio (TBR). Right: Example cuts through the SF-dose plane for specified values of FAZA TBR.

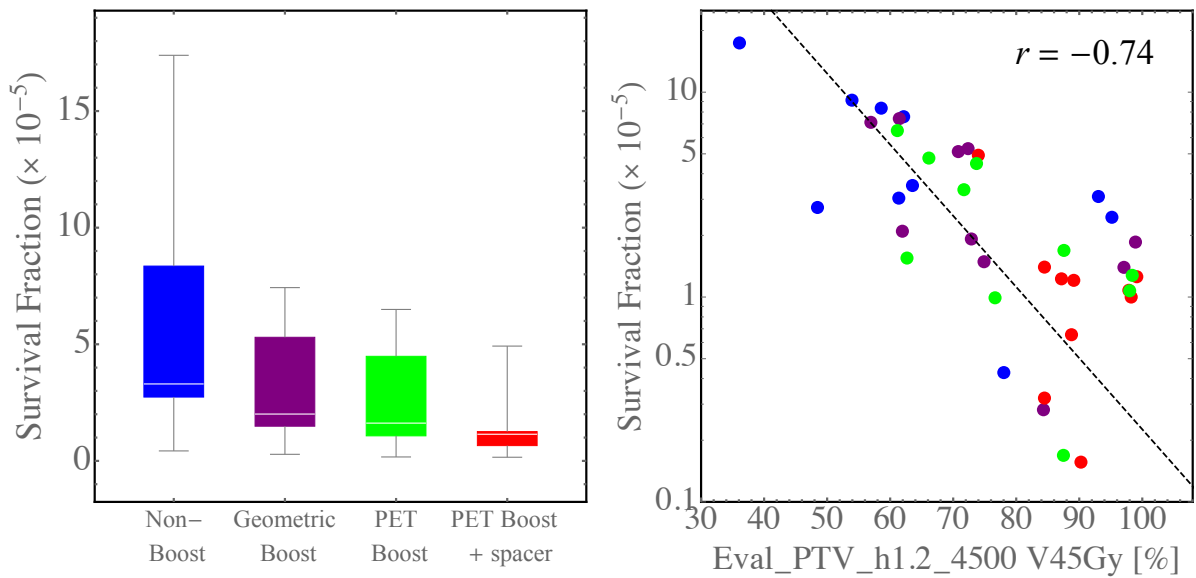

Figure 3. Estimated clonogen cell survival fraction. Left: Box-and-whisker plots for each planning strategy. Right: The fraction of the evaluation PTV corresponding to the lowest hypoxic sub-volume receiving 45Gy predicted survival fraction for all plan strategies ($r$ = -0.74).

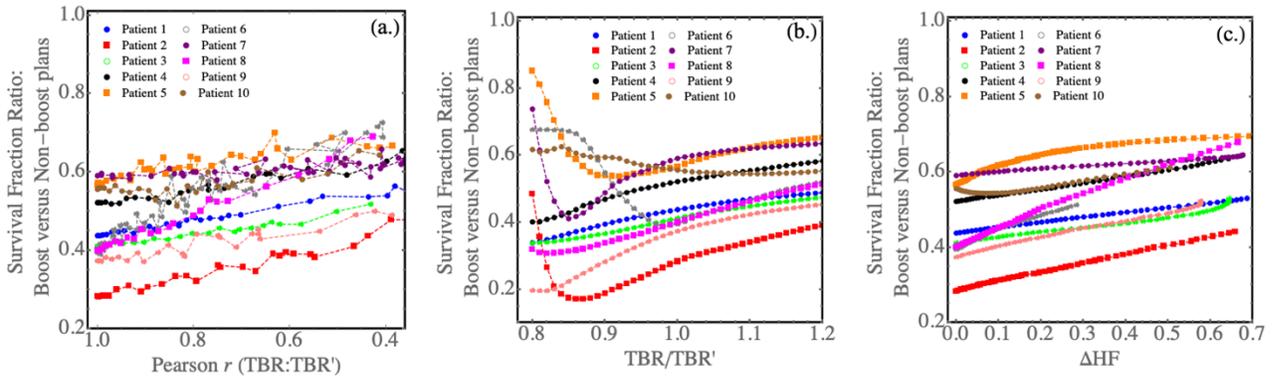

Figure 4. Sensitivity of dose painting plans for all patients to: (a) random noise added to the TBR map, (b) reference tissue quantification errors, and (c) simulated transport effects resulting in an error ΔHF in hypoxic fraction.

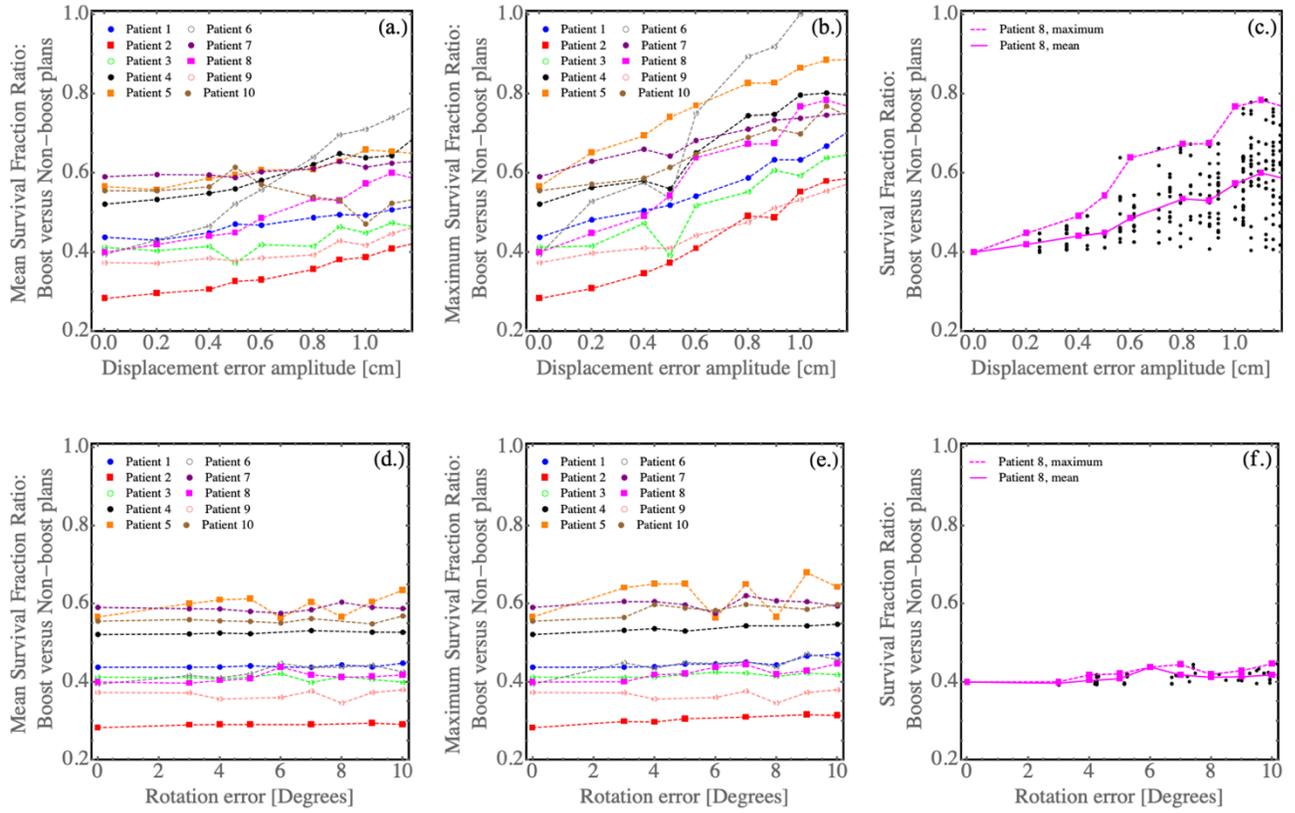

Figure 5. Sensitivity of dose painting plans to registration errors between PET and planning image sets. Top row: ratios of dose painting and standard SBRT SFs for each patient as functions of the amplitude of the displacement error; bottom row: same ratios as functions of the Euler angle ("Rotation error") of rotations. Because there are multiple displacements and rotations that yield the same amplitudes and rotation errors (e.g., the data points in (c.) and (f.) represent individual translations and rotations for a representative patient TBR map), mean and max values of the survival fraction ratio for each displacement amplitude/rotation error are shown in (a.)/(d.) and (b.)/(e.), respectively. (c.) and (f.) show the calculation of the mean (solid lines) and maximum (dashed lines) values for a representative patient.

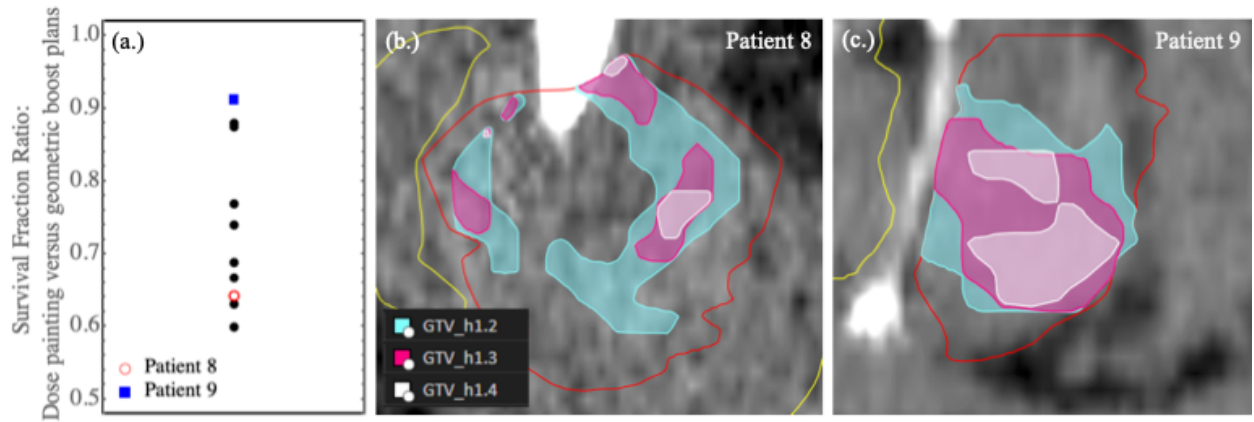

Figure 6. (a.) Distribution of the ratios $SF_{dose\ painting}/SF_{geometric\ boost}$ of *dose painting* and geometric boost plan survival fractions; a smaller value (< 1) indicates that the dose painting plan outperforms the geometric boost plan. (b.) and (c.) Example coronal images of the GTV (red lines) and hypoxic sub-volumes (shaded regions) showing peripheral (patient 8 $SF_{dose\ painting}/SF_{geometric\ boost}$ = 0.64) and central (patient 9, $SF_{dose\ painting}/SF_{geometric\ boost}$ = 0.91) hypoxia distributions. The former benefit more from PET-based dose painting. Duodenum contours are shown by yellow.

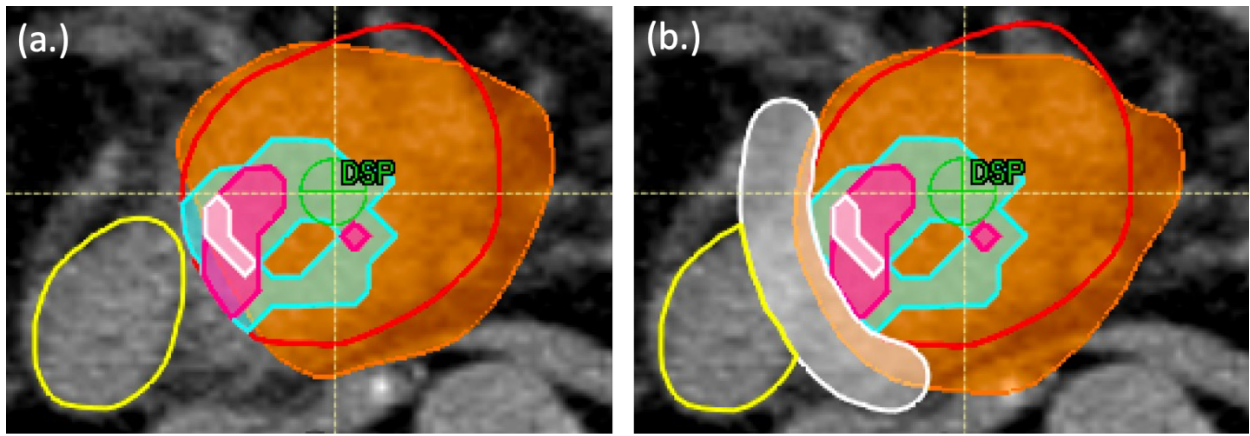

Figure 7. Impact of duodenal spacer on ability to cover the GTV and hypoxic sub-volumes. Axial slice images of the PET boost plan without (a.) and with (b.) simulated duodenal spacer (white filled region) for patient 5: GTV (red lines) shown with hypoxic sub-volumes; the duodenum is shown by yellow lines. The region receiving more than 45 Gy is shown by orange colour wash. Coverage improvement resulted in a reduction of estimated cell survival from $1.6 \times 10^{-5}$ to $3.2 \times 10^{-6}$.